\documentclass[preprint2]{aastex}

\newcommand{\myemail}{viv.maddali@gmail.com}

\newcommand{\fr}{$\Delta \nu_0$}
\newcommand{\frdot}{$\Delta \dot \nu_0$}
\newcommand{\td}{$\tau_d$}
\newcommand{\nupdddot}{\nu_p^{\! \! \! \! \cdots}}

\shorttitle{RXTE HEXTE Analysis of Crab Pulsar Glitch}
\shortauthors{Vivekanand}

\begin{document}

\title{RXTE/HEXTE Analysis of the Crab Pulsar Glitch of July 2000}

\received{15 Jan 2015}
\accepted{by ApJ on 28 Mar 2015}

\author{M. Vivekanand\altaffilmark{1}}
\affil{(Accepted for publication by ApJ on 28 Mar 2015)}

\altaffiltext{1}{No. 24, NTI Layout 1\textsuperscript{st} Stage, 3\textsuperscript{rd} Main, 
1\textsuperscript{st} Cross, Nagasettyhalli, Bangalore 560094, India. (\myemail)}

\begin{abstract}
Hard xray data from the RXTE observatory (HEXTE energy range 15 to 240 keV) have been analyzed 
to obtain a phase coherent timing solution for the Crab pulsar glitch of 15 July 2000. The 
results are: (1) step change in the rotation frequency $\nu_0$ of the Crab pulsar at the 
epoch of the glitch is \fr \ $ = (30 \pm 3) \times 10^{-9} \times \nu_0$, (2) step 
change in its time derivative is \frdot\ = $ (4.8 \pm 0.6) \times 10^{-3} \times \dot \nu_0$, 
and (3) the time scale of decay of the the step change is \td\ $= 4.7 \pm 0.5$ days. The 
first two results are consistent with those obtained at radio frequencies by the Jodrell 
Bank observatory. The last result has not been quoted in the literature, but could be an 
underestimate due to lack of observations very close to the glitch epoch. By comparing with 
the monthly timing ephemeris published by the Jodrell group for the Crab pulsar, the time 
delay between the {main peaks of the} hard xray and radio pulse profiles is estimated to be 
$+411 \pm 167$ $\mu$sec. Although this number is not very significant, it is consistent with 
the number derived for the 2 to 16 keV energy range, using the PCA instrument of RXTE. 
{The separation between the two peaks of the integrated pulse profile of the Crab 
pulsar, and the ratio of their intensities, both are statistically similar before and 
after the glitch. The dead time corrected integrated photon flux within the integrated 
pulse profile appears to decrease after the glitch, although this is not a statistically 
strong result.} This work achieves what can be considered to be almost absolute timing 
analysis of the Crab pulsar hard xray data.

\end{abstract}


\section{Introduction} \label{sec1}

Glitches in pulsars are events in which the rotation period and its derivative (and possibly higher 
derivatives as well) undergo an abrupt change in value, on time scales of less than minutes, often 
followed by a recovery to approximately the pre glitch values, over time scales of days to tens of 
days, or even much longer; see \citep{Shemar1996, Lyne2000, Wong2001, Espinoza2011} for details of 
pulsar glitches, their history and their relevance. Glitches are important to study because they are 
probably one of the very few methods available to study the internal structure of neutron stars 
\citep{Baym1969}; see also \citep{Ruderman1998} and references therein. Glitches are rare events; 
in the Crab pulsar (PSR B0531+21 or J0534+2200) they occur once in $\approx 1.6$ years 
\citep{Espinoza2011}. A typical glitch in the Crab pulsar involves a very small fractional change 
of rotation period (or alternately, rotation frequency) of about $10^{-7}$ to $10^{-9}$ 
\citep{Wong2001, Espinoza2011}. Coupled with the sudden 
onset of a glitch, this implies that frequent and regular pulsar timing observations are required 
to study the glitch phenomenon. The very low rotation period of the Crab pulsar ($P \approx 33.5$ 
ms, or alternately very high rotation frequency $\nu \approx 29.851$ Hz, at the middle of the year 
2000) necessitates timing observations at least twice a day, over a period of one year, to properly 
analyze a glitch; see \citep{Manchester1977, Backer1986, Lyne2006} for pedagogical reviews of 
pulsar timing in general, and analysis of pulsar glitches in particular.

Clearly a dedicated telescope is required to study pulsar glitches. At radio frequencies 
this has been done (and continues to be done) for the Crab pulsar by the Jodrell Bank 
observatory \citep{Lyne1993}. Over the last $\approx 40$ years, they have accumulated 
pulse timing information of the Crab pulsar at 610 and 1400 MHz radio frequencies, 
observing it daily, and have published (and continue to update) the monthly timing 
ephemeris\footnote{{http://www.jb.man.ac.uk/pulsar/crab.html}} of the Crab pulsar; see 
\citep{Lyne1993} for details; {also see \citep{Lyne2015} for the 45 year rotation history 
of the Crab pulsar.} The Crab pulsar has also been observed daily by the Green Bank 
Telescope, at 327 and 610 MHz radio frequencies \citep{Backer2000, Wong2001}. {Ideally 
such work should have been carried out at xray energies, which are not affected by 
problems associated with propagation through the interstellar medium, that radio 
signals are susceptible to \citep{Lyne1993, Backer2000}. However, xray telescopes are 
difficult to build and expensive, in comparison to radio telescopes.} RXTE is one of 
the few xray observatories that can time the arrival 
of pulses from pulsars to the accuracy required for timing analysis. However RXTE is not a 
dedicated pulsar timing observatory, and most Crab pulsar observations of RXTE are spaced, 
on the average, two weeks apart. Fortunately, during the period late 1999 to late 2000, 
three sets of very closely spaced RXTE observations of the Crab pulsar were available, one 
of them being just after glitch of 15 July 2000. It is mainly these three clusters of
observations, and the existence of observations immediately after the glitch, that have 
motivated this work.

This work is organized as follows. Section~\ref{sec2} and section~\ref{sec3} describe HEXTE 
data and the method of analysis. Section~\ref{sec4} presents phase coherent timing results 
for the July 2000 glitch of the Crab pulsar. Section~\ref{sec5} contains discussion and 
describes the behavior of some Crab pulsar parameters before and after the glitch. 

\section{Observations} \label{sec2}

The HEXTE instrument \citep{Rothschild1998} of RXTE consists of two independent clusters of 
detectors, labeled clusters 0 and 1. Each cluster contains four NaI(Tl)/CsI(Na) phoswich 
scintillation photon counters, and has a field of view of one degree in the sky. For 
practical purposes this instrument is sensitive to photons in the 15 to 240 keV range, and 
each photon's arrival time is measured with an accuracy of $\approx 7.6$ $\mu$sec (see 
``The ABC of XTE'' guide on the RXTE 
website\footnote{heasarc.gsfc.nasa.gov/docs/xte/data\_analysis.html}). During normal 
operation, the two clusters switch between the source and a background region of the sky, 
such that when one cluster is pointed at the source, the other is pointed at the background, 
and vice versa. For pulsar observations another mode of observation is also used, in which 
both clusters dwell only on the source. The data used in this work consist of both modes, 
but predominantly of the latter kind. 

The first observation used in this work was obtained on 18 Dec 1999, and the last on 24 Dec 
2000; the corresponding observation identification numbers (ObsID) for the data are 
40090-01-01-00 and 50804-01-14-00, respectively. The epoch of the glitch is MJD $51740.656 
\pm 0.002$ \citep{Espinoza2011}, which is at $\approx$ 15:45 UTC on 15 July 2000. The pre 
glitch data is relatively more frequently observed during the months Dec 1999 and Jan 2000, 
but not later. It extends up to {14 May 2000} only (ObsID 50099-01-26-00), which implies that 
no observations exist for the two month duration just prior to the glitch. Fortuitously, the 
first post glitch observation is on 17 July 2000 at $\approx$ 00:45 UTC (ObsID 50098-01-01-00), 
which is just $\approx 1.4$ days after the glitch. From then onward the data is well sampled 
(frequently observed) up to 31 Jul 2000 (ObsID 50099-01-02-00), after which the data is under 
sampled until 5 Dec 2000 (ObsID 50099-01-11-00), which is for most of the post glitch duration.  
Then again the data is well sampled until 24 Dec 2000 (ObsID 50804-01-14-00). There are $50$ 
ObsID during this period, out of which $2$ were not useful; during ObsID 50100-01-01-05 the 
Crab pulsar was completely occulted by the {Earth} (ELV $< 0^\circ$, {which is the instantaneous 
angle between the Earth's limb and the astronomical source}); and ObsID 50099-01-05-00F 
was obtained on 11 Sept 2000, during the week when there were up to $1^\circ$ errors in the 
spacecraft attitude. Six of the remaining 48 ObsIDs have more than $10$\% data 
gaps in them, but they have sufficient useful data for our purpose. Thus the Crab pulsar is 
under sampled (from the timing point of view) for most of the one year duration under 
consideration, except for three brief periods of well sampled data, one at the very beginning 
of the pre glitch duration, one just after the glitch, and one at the very end of the 
observations used in this work.

In fact the pre and post glitch duration were chosen based partly on the availability of 
closely spaced (frequently observed) data. The other reason was for the data to be 
sufficiently isolated in time from the previous and the next Crab pulsar glitches, so that 
the timing analysis is not corrupted by their residual effects. The previous Crab pulsar 
glitch (a small glitch) occurred at $\approx$ 00:29 UTC on 1 Oct 1999, and the next glitch 
(a large one) occurred at $\approx$ 01:44 UTC on 24 Jun 2001 \citep{Espinoza2011}; {the 
very small glitch at $\approx$ 18:00 UTC on 17 Sept 2000 is ignored} because the data of
this work is not sensitive to it. The first observation of this work is 78 days after the 
previous glitch, which is several times the {longest of the short decay timescales for the 
Crab pulsar \citep{Lyne2000, Wong2001, Wang2012}}, {so one expects} that the pre glitch 
timing solution in this work is not corrupted by the decay phenomenon of the previous 
glitch. The last observation of this work is six months before the next big Crab pulsar 
glitch. 

\section{Data Processing} \label{sec3}

All data used in this work have been acquired in the \textbf{Event List} mode (operating 
modes E\_8us\_256\_DX0F, E\_8us\_256\_DX1F, etc.), in which the photon arrival times have 
accuracy $\approx 7.6$ $\mu$sec (the best possible for HEXTE), and also the highest energy 
resolution (256 channels in the energy range 0 to 250 keV); see ``Reduction and Analysis 
of HEXTE data" on the RXTE 
website\footnote{heasarc.gsfc.nasa.gov/docs/xte/recipes/cook\_book.html}. Throughout this 
analysis, data of each cluster are analyzed separately.

The first step in data processing is the creation of the so called Good Time Intervals 
(GTI), which are time duration identifying the useful data, not corrupted by instrumental 
and extraneous factors. The GTI that account for system provided checks are created by 
the tool \textbf{maketime} using the filter file available in the directory STDPROD, and 
using the screening criterion (1) ELV $> 10^\circ$, (2) {the difference between the source 
position and the pointing of the satellite} (OFFSET) $< 0.02^\circ$, (3) {the time since the 
peak of the last South Atlantic Anomaly passage} (TIME\_SINCE\_SAA) $> 30$ min or 
TIME\_SINCE\_SAA $< 0$ min.

\subsection{Standard Processing of HEXTE Data} \label{sec31}

The next step is to create additional GTI based on the light curves of the observation; 
for example, abrupt changes of large magnitude in photon count rates should be excluded 
from the analysis.

For that, each data file for each cluster is processed by the tool \textbf{hxtback} to 
separate the data pertaining to source and sky background regions; for timing analysis only 
the source data are used. Then light curves are obtained using the tool \textbf{seextrct}, 
screening the data using the GTI available in each file, as well as the GTI file created 
in the previous section. The light curves are binned at the telemetry interval DELTAT (16 
sec), and photons are selected from energy channels 15 to 240 keV. The light curves are 
corrected for dead time using the tool \textbf{hxtdead}, using the appropriate house keeping 
file. Although all four detectors are chosen for both clusters while running the 
{tools}, the third detector of cluster 1 lost ability to assign energy 
information to a photon after 6 March 1996; photons of this detector fall in the first two 
energy channels irrespective of their actual energy. Therefore our choice of 15 to 240 keV 
energy range essentially filters out photons from this detector, even though they have 
valid arrival time information; see ``The XTE Technical Appendix'' on RXTE 
website\footnote{heasarc.gsfc.nasa.gov/docs/xte/appendix\_f.html}.

Light curves for both clusters are plotted, and the range of count rates, within which the 
data appears good, are chosen. These light curves and count rate limits are used to create 
a second set of GTI files, one for each cluster, using the tool \textbf{maketime}. These 
GTI are then merged with the earlier GTI, using the tool \textbf{mgtime} with the AND 
option, to yield the final GTI, which are then used to filter the individual data files, 
using the tool \textbf{fselect}. By this stage, one has screened the data for all system 
provided checks, as well as for user provided count rate limits. Finally, the photon arrival 
times are referred to the solar system barycenter using the tool \textbf{faxbary}, using the
orbit file for the given ObsID, and the Crab pulsar's coordinates ($83.6332208^\circ$ for 
right ascension and $22.0144611^\circ$ for declination, for the epoch J2000, 
{\citep{McNamara1971}}, taken from the online pulsar catalog of 
ATNF\footnote{www.atnf.csiro.au/research/pulsar/psrcat/}).

\subsection{Obtaining the Period of Crab Pulsar} \label{sec32}

The next step is to obtain the best period of the Crab pulsar for each data file of each
cluster. The fundamental frequency in the power spectrum of the data gives the first 
approximation to the period, which is obtained using the tool \textbf{powspec}, with bin 
size $0.67$ ms and data length $2^{20} = 1048576$ bins for most files, but half or 
quarter of that for shorter files. This is done for each data file of each cluster, not
only to check for consistency of the period among all files of a single ObsID, but also 
to check the health of the data. The second approximation to the period is obtained by 
folding the data over a range of 600 periods centered on the first approximation period, 
and searching for the maximum $\chi^2$,  using the tool \textbf{efsearch}, with an 
increment of $10^{-8}$ sec in period, and zero period derivative. The third approximation 
to the period is obtained by doing a finer search centered on the second approximation
period, using an increment of $0.2 \times 10^{-8}$ sec in period, and a nominal period 
derivative of $420 \times 10^{-15}$ sec per sec {(explained later)}, over a range of 128 
periods. A Gaussian is fit to the $\chi^2$ as a function of period, to obtain the 
centroid. The final period is obtained by folding the earlier and later portions of 
the data at the third approximation period, then cross correlating the two integrated 
profiles, then measuring 
the shift (if at all) versus time between the two profiles. Folding is done using the 
tool \textbf{efold}, while cross correlation is done using independently developed 
software. The accuracy of the final period was typically $7$ nano sec; for ObsID with 
long duration observations it could be as small as fraction of a nano sec.

\subsection{Obtaining the Epoch of the Main peak of Crab Pulsar} \label{sec33}

The next step is to combine data files of each ObsID to get the integrated profile of 
the Crab pulsar separately for each cluster, with a resolution of 128 bins per period, 
using the best period of the epoch, and a nominal period derivative of $420 \times 
10^{-15}$ sec per sec, {that is obtained by fitting the periods versus epochs for the
48 ObsIDs; this value is also consistent with the mean period derivative of the Crab 
pulsar for the duration Dec 1999 to Dec 2000, as estimated from radio data, given in 
the Jodrell monthly ephemeris.}
The zero phase of the integrated profile is set to the 
arrival time of the first valid photon in the data, after conversion to MJD by adding 
the MJDREF available in the data files.  For validity of this procedure, it must be 
ensured that the keyword RADECSYS in the barycenter corrected data files is set to FK5, 
the keyword TIMEZERO is set to $0$, and the keyword CLOCKAPP is set to T; see ``A Time 
Tutorial" in ``The ABC of XTE'' guide, and also ``RXTE Absolute Timing Accuracy" on RXTE 
website\footnote{heasarc.gsfc.nasa.gov/docs/xte/abc/time.html}.

\begin{figure}[h]
\epsscale{1.0}
\plotone{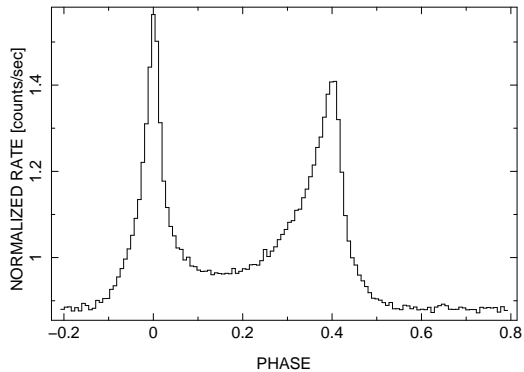}
\caption
{\small
Integrated profile of Crab pulsar obtained by combining data of ObsID 
40090-01-01-00 from both clusters. The epoch of the peak of the main pulse of 
cluster 0 data is used as reference phase for the tool \textbf{efold}, and 
data is folded at the period of this epoch (0.0335046788 sec). The average 
counts per sec (used to normalize the ordinate) is 176.477783. \label{fig1}
}
\end{figure}

The final step is to obtain the epoch of arrival of the peak of the main pulse of Crab 
pulsar, which is taken as the fiducial point in its integrated profile. For this the 
epoch of the zero phase of the integrated profile (described above) should be added to 
the position of the peak of the main pulse. This position is found by three independent 
means, as done in \citep{Rots2004} but with some difference -- (1) fitting a Gaussian 
to the main peak data, (2), fitting a 
Lorentzian to the main peak data, and (3) finding the first moment of the main peak 
data higher than $80$\% of the peak value. The first difference with \citep{Rots2004} 
is that they fit a parabola instead of a Gaussian in method (1). The second difference 
is that they have 200, 400 and 800 bins in the integrated profile for the three methods, 
respectively, whereas the number of bins in this work are 128 for all three methods, 
because several ObsID do not have sufficient exposure to obtain sufficient signal to 
noise ratio with higher number of bins. The disadvantage of having lower number of 
bins in the integrated profile is that one has to be cautious while fitting the 
Gaussian and the Lorentzian; one has to choose as much of the main peak data as 
possible, to maximize the sensitivity of the fit, but should not include the asymmetric 
parts of the peak in its wings. {This fitting is done for each cluster for each ObsID.}
All three methods give consistent timing results for 
the Crab pulsar glitch of July 2000. The typical accuracy of the fit is $\approx 1$ 
milli period. Figure ~\ref{fig1} shows the integrated profile of the Crab pulsar for 
ObsID 40090-01-01-00 for the combined data of both clusters.

\section{Phase Coherent Timing Solution for the July 2000 Glitch} \label{sec4}

The timing solution for Crab pulsar presented in this work was obtained using the 
Gaussian fitting method described earlier, with three variations of epoch of 
arrival of the peak of the main pulse (henceforth referred to as pulse arrival epoch)
-- (1) epochs from data of cluster 0 only, (2) epochs from data of both clusters, but 
separately, and (3) epochs from combined data of both clusters. Methods (1) and (3) 
both have 48 epochs, while method (2) has twice the number. The epochs from method 
(3) are expected to be the most reliable due to enhanced number of photons in the 
integrated profile by the average factor 1.75 (cluster 1 gathers $25$\% less photons 
due to excluding detector 3). All three methods give consistently similar results. 
The results presented in this work are derived using method (3). As mentioned earlier,
timing solutions obtained using method (3) along with fitting a Lorentzian, and 
finding the first moment, all give results similar to those presented in 
Table~\ref{tbl2} in this work. In addition, compared to the Gaussian fit method, the 
first moment method gave a mean departure of $1 \pm 2$ milli periods for the 48 
epochs; for the Lorentzian method the mean departure was $0.3 \pm 0.4$ milli periods. 
Thus it is concluded that pulse arrival epochs are consistent among the three peak 
finding methods.

\begin{figure}[h]
\epsscale{1.0}
\plotone{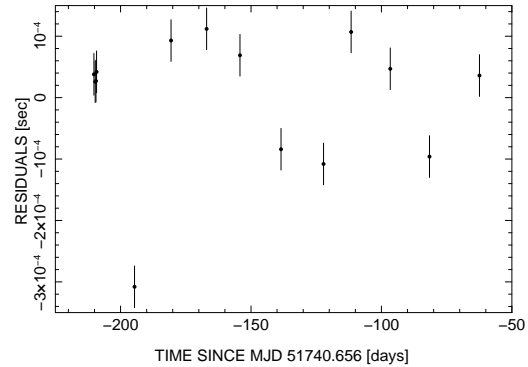}
\caption
{\small
Result of using TEMPO2 on the 14 pre glitch pulse arrival epochs, fitting for $\nu$, 
$\dot \nu$ and $\ddot \nu$, using the epoch of the first data at -210.221 (MJD 
51530.4349129857) as the reference epoch for phase zero in TEMPO2; the results are 
given in Table~\ref{tbl1}. The origin of the abscissa is at the epoch of the glitch 
(MJD 51740.656). Note the four closely spaced epochs at the start of the data, and
the lack of any observation for two months prior to the glitch. \label{fig2}
}
\end{figure}

\begin{table}[h]
\begin{center}
\caption{\small TEMPO2 best fit parameters to the pre glitch data of Figure~\ref{fig2}. 
$\nu$ is the rotation frequency of the Crab pulsar at the epoch MJD 51530.4349129857, 
which is the first epoch in our data; $\dot \nu$ and $\ddot \nu$ are the first and 
second time derivatives of $\nu$, respectively, at the same epoch. The errors in 
brackets are in the last digit of each result. \label{tbl1}}
\begin{tabular}{|l|l|}
\tableline
Parameter  & Value \\
\tableline
$\nu$ (Hz)  & $29.846592902(2)$  \\
\tableline
$\dot \nu$  ($10^{-10}$ s$^{-2}$) & $-3.745962(8)$ \\
\tableline
$\ddot \nu$  ($10^{-20}$ s$^{-3}$) & $0.94(1)$ \\
\tableline
\end{tabular}
\end{center}
\end{table}

Figure ~\ref{fig2} shows the result of using TEMPO2 \citep{Hobbs2006} on the 14 pre
glitch pulse arrival epochs. {In the absence of glitches and timing noise, pulsars 
obey a simple slowdown model that is adequately represented, at least for short 
duration of several months, by three parameters -- rotation frequency $\nu$, its 
time derivative $\dot \nu$ and its second derivative $\ddot \nu$; see Equation 1 of
\citep{Espinoza2011}. The three fitted parameters are shown in Table~\ref{tbl1},}
and are consistent with the interpolated values from the Jodrell monthly ephemeris 
for the Crab pulsar for that epoch. The formal one standard deviation 
error on the pulse arrival epoch in Figure~\ref{fig2} is typically $0.87$ milli 
periods. An independent estimate of the error is obtained by the difference in 
pulse arrival epochs for clusters 0 and 1 for the same ObsID.  Often these differ 
by $\approx 16$ sec or a few multiples of it, but on rare occasions can differ 
by hours. Ideally these relatively short duration differences should be equivalent 
to almost integer number of pulse cycles, since the estimated period of rotation
of the Crab pulsar would be very accurate for closely spaced epochs; the departure 
from integer values will give us an idea of the errors involved in pulse arrival 
epochs in Figure~\ref{fig2}. The mean value of the departure from integer number 
of cycles turns out to be $\approx 1.0$ milli period. Therefore all error bars in 
Figure ~\ref{fig2} are set to one milli period. The rms residual of the 14 pre 
glitch arrival times after the TEMPO2 fit is $3.2$ milli periods.

\begin{figure}[h]
\epsscale{1.0}
\plotone{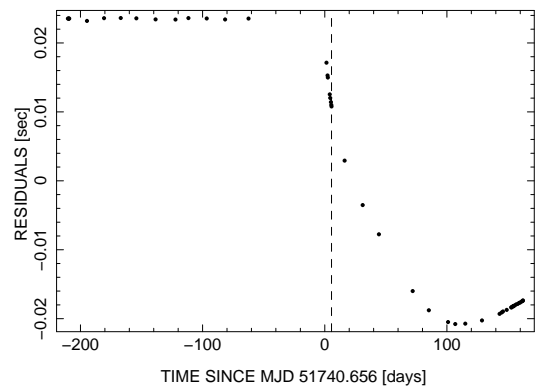}
\caption
{\small
Result of using TEMPO2 on the 34 post glitch pulse arrival epochs, using the values
of Table~\ref{tbl1} as input parameters and without fitting. The origin of the abscissa 
is at the epoch of the glitch (MJD 51740.656). The vertical dashed line is at epoch 
after which one phase cycle was subtracted for all subsequent epochs, using the command 
PHASE $-1$ in the input file to TEMPO2. Note the clusters of closely spaced epochs, one 
just after the glitch epoch, and the other at the end of the data. \label{fig3}
}
\end{figure}

TEMPO2 is used with the parameters of Table~\ref{tbl1} as constant input (i.e., 
without any fitting) for the 34 post glitch pulse arrival epochs; the results are 
shown in Figure~\ref{fig3}. These post glitch residuals $\Delta \phi$ (in sec) are 
fit to a {modified version} of the glitch model of \citep{Shemar1996},


\begin{eqnarray}
\Delta \phi & = & -\frac{1}{\nu_0} \int_{\epsilon}^t dt \left [ \Delta \nu_p + 
     \Delta \dot \nu_p t + \Delta \nu_n \exp \left ( -\frac{t}{\tau_d} \right ) 
     \right ] \nonumber \\
            & \approx & -\Delta \phi_0 - \frac{\Delta \nu_p}{\nu_0} t 
              - \frac{\Delta \dot \nu_p}{\nu_0} \frac{t^2}{2}  \nonumber \\
            & &   - \frac{\tau_d \Delta \nu_n}{\nu_0} 
            \left ( 1 - \exp \left ( - \frac{t}{\tau_d} \right ) \right ),
\end{eqnarray}

\noindent
where $t$ {is the time elapsed since the glitch epoch} (in seconds), $\nu_0$ is the 
rotation frequency of the Crab pulsar at the epoch of the glitch (in Hz), $\Delta 
\nu_p$ and $\Delta \nu_n$ are the permanent and exponentially decaying parts of 
the step change in rotation frequency at the epoch of the glitch, $\Delta \dot 
\nu_p$ is the permanent step change in the time  derivative of the rotation 
frequency, and $\tau_d$ is the decay time scale (in sec) of the step frequency 
change. {The parameter $\Delta \phi_0$ accounts for any uncertainty 
$\epsilon$ in the epoch of the glitch, which is assumed to be much smaller 
than $\tau_d$.
The equation above differs from Equation 1 of \citep{Shemar1996} in having a single 
exponential only; as mentioned in their paper, only occasionally one requires more 
than one transient component, and the decay times of these additional transients 
are typically hundreds of days.} The negative sign in Equation 1 accounts for the 
fact that the phase residuals after a glitch increase in the negative direction 
for a positive step change in rotation frequency \citep{Shemar1996}.  
Table~\ref{tbl2} gives the minimum $\chi^2$ fit values of the parameters in 
Equation 1. The rms residual of the 34 post glitch arrival times after the fit 
is $8.6$ milli periods.

\begin{table}[h]
\begin{center}
\caption{\small Minimum $\chi^2$ parameters obtained by fitting Equation 1 to the post glitch 
data of Figure~\ref{fig3}. \label{tbl2}}
\begin{tabular}{|l|l|}
\tableline
Parameter  & Value \\
\tableline
$\Delta \phi_0$ (ms) & $3.3 \pm 0.5$  \\
\tableline
$\Delta \nu_p$  ($10^{-6}$ Hz) & $0.180 \pm 0.003$ \\
\tableline
$\Delta \dot \nu_p$  ($10^{-13}$ s$^{-2}$) & $-0.350 \pm 0.006$ \\
\tableline
$\Delta \nu_n$  ($10^{-6}$ Hz) & $0.71 \pm 0.08$ \\
\tableline
$\tau_d$  (days) & $4.7 \pm 0.5$ \\
\tableline
\end{tabular}
\end{center}
\end{table}

The epoch of the glitch is set to the published value of MJD $51740.656$, since 
our data is unable to verify this number. \citep{Espinoza2011} obtain this by 
using trial glitch epochs as reference epochs in TEMPO2 for the pre and post 
glitch timing solutions, then comparing the two solutions for a match in phase 
at the reference epoch (glitch epoch). This is not possible in this work because 
of under sampling of data; the one and only pre glitch solution available 
(Table~\ref{tbl1}) was with {the reference epoch used}, because close to that the 
data was well sampled. {An alternate method of formulating the parameter 
$\Delta \phi_0$ is to take the integral in Equation 1 from the limits $0$ to 
$t - \epsilon$, instead of from $\epsilon$ to $t$. It can be shown that the two 
methods are equivalent as long as 
$\epsilon / \tau_d \ll 1$, and the derived parameters $\Delta \nu_p$, $\Delta 
\dot \nu_p$ and $\Delta \nu_n$ have the relative orders of magnitude as derived 
in Table 2.}

The total step change in rotation frequency at the glitch, as a fraction of the 
pre glitch frequency, is $(\Delta \nu_p + \Delta \nu_n) / \nu_0 = (30 \pm 3) 
\times 10^{-9}$, {which is not too different from} the value of $(25.1 \pm 0.3) 
\times 10^{-9}$ published by \citep{Espinoza2011}. The actual step change of $\Delta \nu_p + 
\Delta \nu_n \approx 0.89 \pm 0.08$ $\mu$Hz compares well with the top panel of 
Figure 5 of \citep{Espinoza2011}. The fraction of frequency recovery $Q$ is $0.71 
/ (0.71 + 0.18) \approx 0.80 \pm 0.11$, which is very high, as evident from Figure 
5 of \citep{Espinoza2011}, although the decay of frequency in their figure (top 
panel) does not appear to be entirely exponential, {most probably due to the 
very small glitch that has been ignored in this work.}

The total step in frequency derivative at the glitch as a fraction of the pre glitch 
frequency derivative is $(\Delta \dot \nu_p - \Delta \nu_n / (\tau * 86400) ) / \dot 
\nu_0 = (4.8 \pm 0.6) \times 10^{-3}$, {which is also not too different from} the value 
of $(2.9 \pm 0.1) \times 10^{-3}$ of \citep{Espinoza2011}. The fraction of recovery 
of the frequency derivative is $-17.48 / (-17.48 - 0.35) \approx 0.98 \pm 0.16$, 
which is consistent with Figure 5 of \citep{Espinoza2011}. However, the step change 
in $\dot \nu_0 \approx (-18 \pm 2) \times 10^{-13}$ estimated here is inconsistent 
with the bottom panel of Figure 5 of \citep{Espinoza2011}, in which it is more like 
$\approx -5 \times 10^{-13}$.  However, the value expected from their work is 
$\approx 2.9 \times 10^{-3} \times -3.744255 \times 10^{-10} \approx -11 \times 
10^{-13}$, which is in between the above two numbers.

\citep{Espinoza2011} do not quote a value for the decay time scale \td\   for this
glitch. By expanding and gridding their Figure 5, and reading off values of the peak 
and the first decay point, one can obtain an approximate value for \td. This 
turns out to be $\approx 13 \pm 3$ days, for the decay of both the frequency and 
its derivative (top and bottom panels respectively of that figure). The number 
derived in this work, \td\ $= 4.7 \pm 0.5$ days, is a factor of $\approx 
2.8$ smaller, although it is in the right range of decay time scales for Crab pulsar 
\citep{Lyne2000}.  If it turns out that the correct value of \td\  is indeed $\approx 
13$ days, then the reason for the factor of $2.8$ underestimate in this work is most 
probably on account of lack of observations very close to the glitch epoch, to which 
the estimate of \td\  is very sensitive.

\section{Discussion} \label{sec5}

{In section~\ref{sec4} we concluded} that the hard xray timing of the Crab pulsar 
glitch of 15 July 2000, using HEXTE/RXTE data, is consistent with the results 
obtained at radio frequencies by the Jodrell Bank observatory. This is, {to the best 
of our knowledge}, the first time that a glitch of the Crab pulsar has been analyzed 
using xray data, resulting in what can be considered to be the closest to absolute 
timing of the Crab pulsar at hard xray energies.

\begin{figure}[h]
\epsscale{1.0}
\plotone{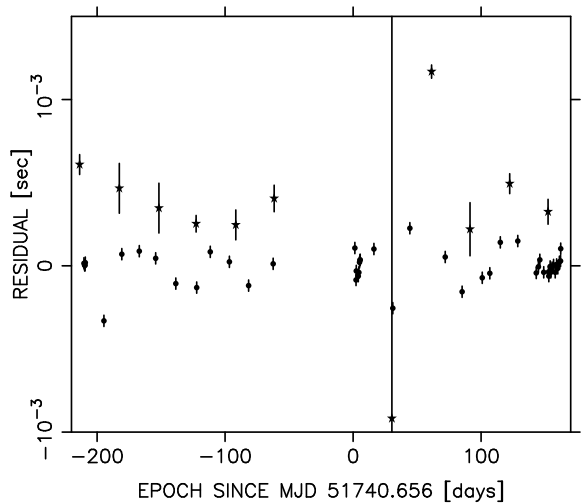}
\caption
{\small
Pre and post glitch xray residuals (this work, dots; {see text for the corresponding 
models}) along with eleven radio residuals from Jodrell Bank data (stars), which lie 
consistently above xray residuals, except for the residual of 15 Aug 2000, which 
has a very large error bar ($4$ ms). \label{fig4}
}
\end{figure}

\citep{Rots2004} found that {the main peak of the xray pulse profile of the Crab 
pulsar (the peak at phase 0 in Figure~\ref{fig1}) leads the main peak of the radio 
pulse profile (just after the radio precursor; see Figure 6a of \citep{Kuiper2003})}
by $344 \pm 40$ $\mu$sec. Figure~\ref{fig4} shows the result of fitting the data of 
Figure~\ref{fig3} combined with the Jodrell data at radio frequencies, obtained 
from their monthly ephemeris. {It shows the combined pre and post glitch residuals 
after subtracting the pre and post glitch models, respectively. The pre glitch 
model is given in Table 1. The post glitch model is obtained by including 
additionally in Equation 1, permanent step changes in the second and third time 
derivatives of the rotation frequency, $\Delta \ddot \nu_p$ and $\Delta \nupdddot$, 
respectively. This solution gave the lowest rms residual of $3.2$ milli periods 
for the 34 post glitch arrival times; the values of the rest of the parameters 
obtained in this fit are consistent with those in Table 2.} The mean pre glitch 
separation of the Jodrell data with respect to the xray data is $411 \pm 
167$ $\mu$sec. Although this result has $\approx 4.2$ times larger error than 
that quoted by \citep{Rots2004}, it is consistent with their result, and also 
with the results of others \citep{Kuiper2003, Carrillo2012}. The post glitch 
differences with Jodrell data are difficult to analyze, due to the baseline 
varying in a quasi periodic manner. However, after accounting for these variations, 
the last three post glitch Jodrell data certainly support the above number.

\begin{figure}[h]
\epsscale{1.0}
\plotone{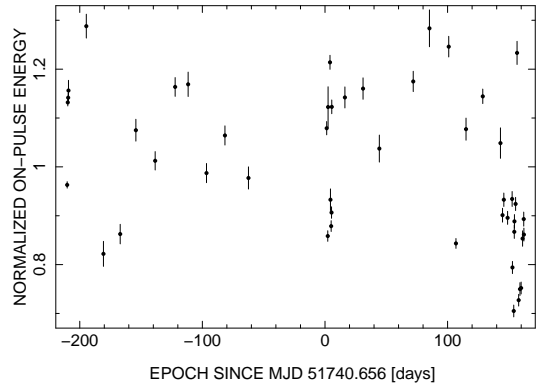}
\caption
{\small
Normalized and dead time corrected on pulse energy of Crab pulsar during the year 
2000. \label{fig5}
}
\end{figure}

Several properties of the Crab pulsar can be studied as a function of pre and post glitch
epochs. Three parameters were plotted as a function of epoch -- (1) The separation of the 
two peaks in the Crab pulsar's integrated profile, (2) the ratio of their peaks, and (3) 
integrated energy in the pulse profile. {The first two showed no variation worth 
reporting} (see also \citep{Rots2004}). Figure~\ref{fig5} shows the dead time corrected 
photon count variation of the Crab pulsar as a function of epoch. 

Each point in Figure~\ref{fig5} is obtained by first estimating the average off pulse 
counts in the integrated profile, then subtracting it from the on pulse counts before 
integrating them, then dividing the result by the average off pulse counts; this is 
done for every DELTAT (16 sec) of data for each ObsID. Off pulse counts are obtained 
by integrating in the phase range $0.61$ to $0.79$ in Figure~\ref{fig1}; the rest of 
phase range represents the on pulse. Although this procedure will exclude any small 
off pulse emission from the Crab pulsar \citep{Tennant2001}, it has the advantage of 
correcting for dead time, which is not available for Barycenter corrected data, and 
also for data that has been filtered in energy range, which is the situation with our 
data; see ``The XTE Technical Appendix'' on RXTE 
website\footnote{heasarc.gsfc.nasa.gov/docs/xte/appendix\_f.html}. Finally all $48$ 
energies are normalized by their mean value, which lies at  the value $1.0$ in 
Figure~\ref{fig5}.  The assumption made here is that dead time correction is, to a 
large order of accuracy, the same for the on and off pulse phases of a pulsar, even 
for bright pulsars such as the Crab. Dead time is the duration immediately after the 
arrival of an xray photon or high energy particle, during which the HEXTE detectors 
are unable to process any more photons. For the HEXTE detector there are two sources 
of dead time. One is due to the arrival of an xray photon, after which the HEXTE 
detectors are "dead" for $16$ to $30$ $\mu$sec.  The much larger effect is due to 
arrival of high energy particles, after which the detectors are dead for $2500$ 
$\mu$sec; these are known as XULD events. Even for a luminous pulsar such as the 
Crab and its nebula, the normal photon count rate is typically $< 400$ photons per 
sec per cluster, while the normal XULD event rate is typically $150$ particles per 
sec per detector (there are $4$ detectors per cluster). Taking the mid value of 
$23$ $\mu$sec, it can be easily seen that the ratio of the photon to XULD 
contribution to the dead time is $\le 0.6$\%. Since the XULD event rate is the 
same for on and off pulse regions, the above method of dead time correction is 
justified. If required, more refined dead time correction can be done by using the 
integrated pulse profile of the pulsar.

In Figure~\ref{fig5}, the standard deviation of the spread in values is $\approx 
16$\% of the mean value; the corresponding spread in the uncorrected fluxes is 
$\approx 49$\% of the mean value. This technique of dead time correction for 
pulsars significantly improves the precision of the estimate of integrated xray 
flux, and may therefore prove useful to find out if the Crab pulsar xray flux 
varies on time scales of decades; for the crab nebula this has already been 
observed at several xray energies (see \citep{Wilson-Hodge2011} and references 
therein).

{In Figure~\ref{fig5} there is no strong evidence for the xray flux of 
the Crab pulsar to be affected by the glitch}. The mean pre and post glitch xray 
fluxes are $1.06$ and $0.98$, respectively, while the standard error on these 
means is $0.03$. The mean xray fluxes differ by $0.08 \pm 0.04$, {which 
can not be considered a strong result. Moreover, this decrease appears to be mainly 
due to the data well after the glitch in Figure~\ref{fig5}, making it less likely 
to be related to the glitch itself. One instrumental feature that can simulate
such a result is a variation of the average off pulse counts as function of epoch 
in Figure~\ref{fig5}; however this does not appear to be the case.} Although this 
can not be considered a strong result, it may be interesting to speculate on the 
possible causes 
of glitch related xray flux variations in rotation powered pulsars. Such studies 
have not been done so far. Although glitch related xray flux enhancements have 
been reported in some Magnetars (see \citep{Espinoza2011} and references therein), 
it is generally believed that in Magnetars and AXPs, glitches are not always 
associated with xray flux variations \citep{Dib2008}. For the Crab pulsar glitch
under consideration here, the post glitch permanent changes in rotation frequency 
$\Delta \nu_p$ and its derivative $\Delta \dot \nu_p$ can contribute at most to 
a fractional increase of $\approx 0.01$\% in the post glitch xray flux, while 
Figure~\ref{fig5} shows a decrease of $8$\%; so simple energy loss rate of the
Crab pulsar may not be the explanation. The glitch model of \citep{Ruderman2009} 
provides a causal connection between a glitch in a rotation powered pulsar and 
changes in its surface magnetic field. In this model, a glitch is caused by excessive 
stresses in the neutron star crust, that are built up by super fluid vortices
moving outwards due to the spin down of the neutron star, dragging with them
magnetic flux tubes. The crust cracks at the instant of the glitch, and then 
adjusts itself to a new configuration, leading to a corresponding reconfiguration 
of the surface magnetic field, presumably both in terms of its field strength 
as well as in terms of its field line structure (curvature of field lines, 
direction of the opening bundle of field lines at the surface, presence of 
higher magnetic multipoles, etc). There are 
several quantitative uncertainties in the predictions from this model. However 
it may be worth exploring if a small change in surface magnetic field structure 
of the Crab pulsar can lead to a non-linearly large change (decrease in the present
case, but maybe an increase in other glitches) in the post glitch xray flux.

{
It is well known that the dispersion measure of the Crab pulsar varies due to 
radio propagation within the Crab nebula \citep{Lyne1993}, and that the radio 
pulses suffer significant and variable refractive and scattering effects 
within the Crab nebula (see \citep{Backer2000} and references therein). Such 
effects can cause errors in the arrival times of the radio pulses, with 
consequent errors on the derived glitch parameters. The above interstellar 
effects are frequency dependent, and are absent for pulses at xray energies. 
The consistency between the radio and hard xray timing results of this work 
imply that the effect of radio pulse propagating through the Crab 
nebula has not been significant during the glitch of July 2000.
}

{
Glitches are one of the two timing irregularities observed in rotation powered 
pulsars, the other being timing noise, which manifests observationally as 
random wandering of timing residuals. It is currently believed that timing 
noise is due to instabilities in the pulsar magnetosphere \citep{Lyne2010}; 
see also \citep{Arons2009} and references therein for a possible theoretical 
explanation in terms of magnetic reconnection. Now, the radio and xray emitting
regions in the Crab pulsar's magnetosphere are supposed to be identical (the 
outer gaps), except for the radio precursor, which is supposed to arise in the 
polar gap, while for other rotation powered pulsars (eg. Vela) the radio and 
xray emitting regions are supposed to be different; see \citep{Harding2009}
and references therein. Therefore simultaneous radio and xray studies of 
timing noise in the Crab pulsar should show similar timing noise properties (
rms residuals, time scales of wandering, etc), but probably not in Vela like pulsars.
}

{
Glitches, on the other hand, are supposed to be due to the steady differential 
spin down of the super fluid core and outer crust of the neutron star; see
\citep{Ruderman2009} and references therein. At the instant of the glitch, 
the outer crust speeds up, thereby also speeding up the pulse emitting regions 
in the magnetosphere, which are firmly anchored into the crust by means of the
magnetic field. Simultaneous timing observations of pulsar glitches at radio 
and xray wavelengths can probably be used to find out if the emission regions 
at different energies are anchored equally firmly onto the surface of the 
neutron star. For example in the case of the Crab pulsar one would expect the 
glitch behavior to be almost identical, while for the Vela like pulsars one
may or may not notice differences.
}

\acknowledgments

{\small

I thank the anonymous referee for detailed comments to improve this manuscript.
This research made use of data obtained from the High Energy Astrophysics Science 
Archive Research Center Online Service, provided by the NASA-Goddard Space Flight 
Center.
}

{\small

{\it Facilities:} \facility{RXTE (HEXTE)}.
}

\end{document}